\pdfoutput=1
\documentclass[review, 12pt]{elsarticle}
\usepackage{amssymb}
\usepackage{lineno,hyperref}
\usepackage{subfigure}
\usepackage{siunitx, gensymb}
\usepackage{booktabs}
\usepackage{relsize}

\def\textsubscript#1{\ensuremath{_{\mbox{\textscale{.6}{#1}}}}}
\def\textsuperscript#1{\ensuremath{^{\mbox{\textscale{.6}{#1}}}}}
 
\journal{International Journal of Refractory Metals and Hard Materials}

\begin{document}

\begin{frontmatter}

  \title{Effect of ZrB2 additions on the thermal stability of polycrystalline diamond}

\author[add1,add2]{Melisha Jivanji}

\author[add3]{Roy Peter Forbes}

\author[add2]{Humphrey Sithebe}

\author[add1]{Johan Ewald Westraadt\corref{cor1}}
\ead{johan.westraadt@mandela.ac.za}

\cortext[cor1]{Please address correspondence to Johan Ewald Westraadt}
\address[add1]{Centre for HRTEM, Department of Physics, Nelson Mandela University, Gqeberha, 6031, South Africa}
\address[add2]{Element Six (Pty) Ltd, Springs, 1559, South Africa}
\address[add3]{School of Chemistry, University of the Witwatersrand, Johannesburg, 2000, South Africa}

\begin{abstract}
  This study investigates the effect of ZrB\textsubscript{2} additions on the microstructure, thermal stability, and thermo-mechanical wear behaviour of polycrystalline diamond. Following high-pressure high-temperature (HPHT) sintering, the \emph{ZrB\textsubscript{2}-PCD} material showed a full conversion of the binder phase to cobalt-boride (Co\textsubscript{2}B and B\textsubscript{6}Co\textsubscript{23}) phases. In-situ PXRD and TEM vacuum annealing experiments observed that the onset of bulk graphitisation occurred above \SI{1000}{\celsius} for the \emph{ZrB\textsubscript{2}-PCD} material, compared to \SI{850}{\celsius} for the \emph{STD-PCD} material. The \emph{ZrB\textsubscript{2}-PCD} tools showed excellent thermo-mechanical wear behaviour, exhibiting increased durability and a steady wear scar progression during high-temperature dry-VTL testing. However, lowered abrasion wear resistance was observed for the \emph{ZrB\textsubscript{2}-PCD} tools during low-temperature wet-VTL testing, probably due the reduced diamond contiguity in the ZrB\textsubscript{2} additive sample. Further optimisation of the ZrB\textsubscript{2} additive phase content, mixing methodology, or sintering conditions could be explored to improve the abrasive wear resistance of this novel PCD material.  
\end{abstract}

\begin{graphicalabstract}
\end{graphicalabstract}

\begin{highlights}
  \item ZrB\textsubscript{2} additions converted the catalytic cobalt into non-catalytic Co\textsubscript{2}B and Co\textsubscript{23}B\textsubscript{6} phases.
  \item Onset of bulk graphitisation only occurred at temperatures above \SI{1000}{\celsius} for the \emph{ZrB\textsubscript{2}-PCD}.
  \item \emph{ZrB\textsubscript{2}-PCD} had excellent thermo-mechanical wear behaviour compared to the standard PCD.
\end{highlights}


\end{frontmatter}


\section{Introduction}
Conventional polycrystalline diamond (PCD) tools are manufactured by sintering diamond powder on a cemented tungsten carbide (WC-Co) substrate, using high pressure ($>$5.5 GPa) and high temperature ($>$\SI{1450}{\celsius}) conditions. During the sintering process, cobalt from the substrate melts and infiltrates into the diamond layer, facilitating the bonding of the diamond grains into a tough, intergrown diamond network. The resulting material has a high micro hardness (\SI{50}{GPa}) and isotropic fracture toughness \SI{8}{\mega Pa {m^{0.5}}} \citep{cook2000} making it an excellent material in the non-metallic abrasives industries for rock drilling \citep{scott2006}, woodworking \citep{philbin2005}, road-planing, and automotive manufacturing \citep{astakhov2015}. 

Despite its superior room temperature properties, PCD materials suffer from thermal instability when exposed to temperatures in excess of \SI{700}{\celsius} \citep{mehan1989, sneddon1987, tze-pin1992, westraadt2015a}. The residual metallic binder present in the PCD material following sintering, initiates thermal instability of the tool by inducing graphitisation \citep{mehan1989, tze-pin1992, westraadt2015a}, oxidation \citep{evans1962, miess1996}, and thermal expansion effects \citep{cook2000, gu2016, shimada2004} at elevated temperatures. These thermally induced degradation mechanisms could lead to premature tool failure and reduced performance.

In this study, thermal stability is defined as the maximum temperature the material can be exposed to before the immediate onset of thermal degradation. Thermal degradation refers to permanent changes occurring in the material during high temperature exposure, which could include stress-induced cracking, graphitisation, and/or oxidation. There are a variety of methods that can be used to evaluate a materials' thermal stability. The first group of methods involve heat treating the material in a controlled atmosphere and analysing microstructural changes that occur in the material during or after high temperature exposure \citep{akaishi1996, miess1996, westraadt2015a}. This method typically evaluates a specific degradation mechanism (graphitisation, oxidation, or cracking) at controlled temperatures depending on the analysis technique. The second group of methods involve mechanical abrasion testing with a significant thermal component. Increased cutting speeds and/or the absence of a coolant will result in the generation of high temperatures at the tool interface \citep{westraadt2015a}. Since the temperatures at the tool interface are not well controlled, it is usually only used as a ranking tool and the results are usually compared to a suitable benchmark. This method, however, combines all the degradation mechanisms into one test, making it a closer representation of real-world applications. The third method to evaluate thermal stability involves high temperature annealing of the PCD tools in a furnace with a controlled atmosphere and then performing a standard mechanical test thereafter \citep{akaishi1996, westraadt2007}. This method measures the effects of thermal degradation on the mechanical performance of the material. In this study, the thermal stability of the PCD materials were assessed based on its propensity to graphitise during thermal treatments performed under vacuum (method 1) and mechanical abrasion testing using a vertical turning lathe (VTL) with and without a coolant (method 2).

The current industry standard to overcome thermal degradation in PCD materials and prolong the tool life is to remove the metallic cobalt binder phase by acid leaching \citep{bovenkerk1981, griffin2002}.  Alternatively, ultrahigh pressures and non-metallic binder systems have been used to either minimize or eliminate the metallic binder phase from the PCD material. Akaishi and co-workers \citep{akaishi1988} used high pressure (\SI{7.7}{GPa}) and temperature (\SI{2000}{\celsius}) to sinter cobalt-PCD with only 1.5 vol.\% metallic binder. Sumiya and co-workers \citep{sumiya2004a} sintered diamond in the absence of a cobalt binder, employing ultrahigh pressure (\SI{15}{GPa}) and temperature (\SI{2300}{\celsius}) to directly convert graphitic/carbonaceous material into ultrahard nano-polycrystalline diamond. Ceramic binder phases such as TiC \citep{chagas2019, hong1999}, TiB\textsubscript{2} \citep{jaworska2015, sha2020}, and MAX phases \citep{jaworska2001} have also been used as successful sintering aids for PCD. Sintering of PCD using non-metallic sintering aids has also been demonstrated with carbonates \citep{akaishi1996,westraadt2007,qian2012}. Although these experimental studies were successful in sintering PCD materials, these novel approaches either resulted in reduced mechanical properties (toughness, hardness or wear resistance), due to a lack of adequate diamond intergrowth, or required elevated sintering conditions ($>$\SI{7.7} \& $>$\SI{2000}{\celsius}), which significantly increase the costs of tool manufacturing. Consequently, none of these methodologies are currently commercially available in the market.

The present study evaluates the effect of ZrB\textsubscript{2} additives on the thermal stability of standard PCD. The addition of additives to the diamond powder does not require major modifications to the existing tool manufacturing technologies. Thus, there are negligible impacts on the manufacturing costs associated with this approach. The aim of the study is to evaluate the effect of ZrB\textsubscript{2} additions on the thermal stability (onset temperature of graphitisation) of polycrystalline diamond composites measured using powder X-ray diffraction (PXRD) and advanced electron microscopy. Mechanical VTL testing was then used to determine the effect of ZrB\textsubscript{2} on the abrasion performance of the material at low temperature (with coolant) and high temperature (no coolant) conditions.

\section{Methodology}

\begin{figure}
  \begin{center}
    \includegraphics[width=\textwidth]{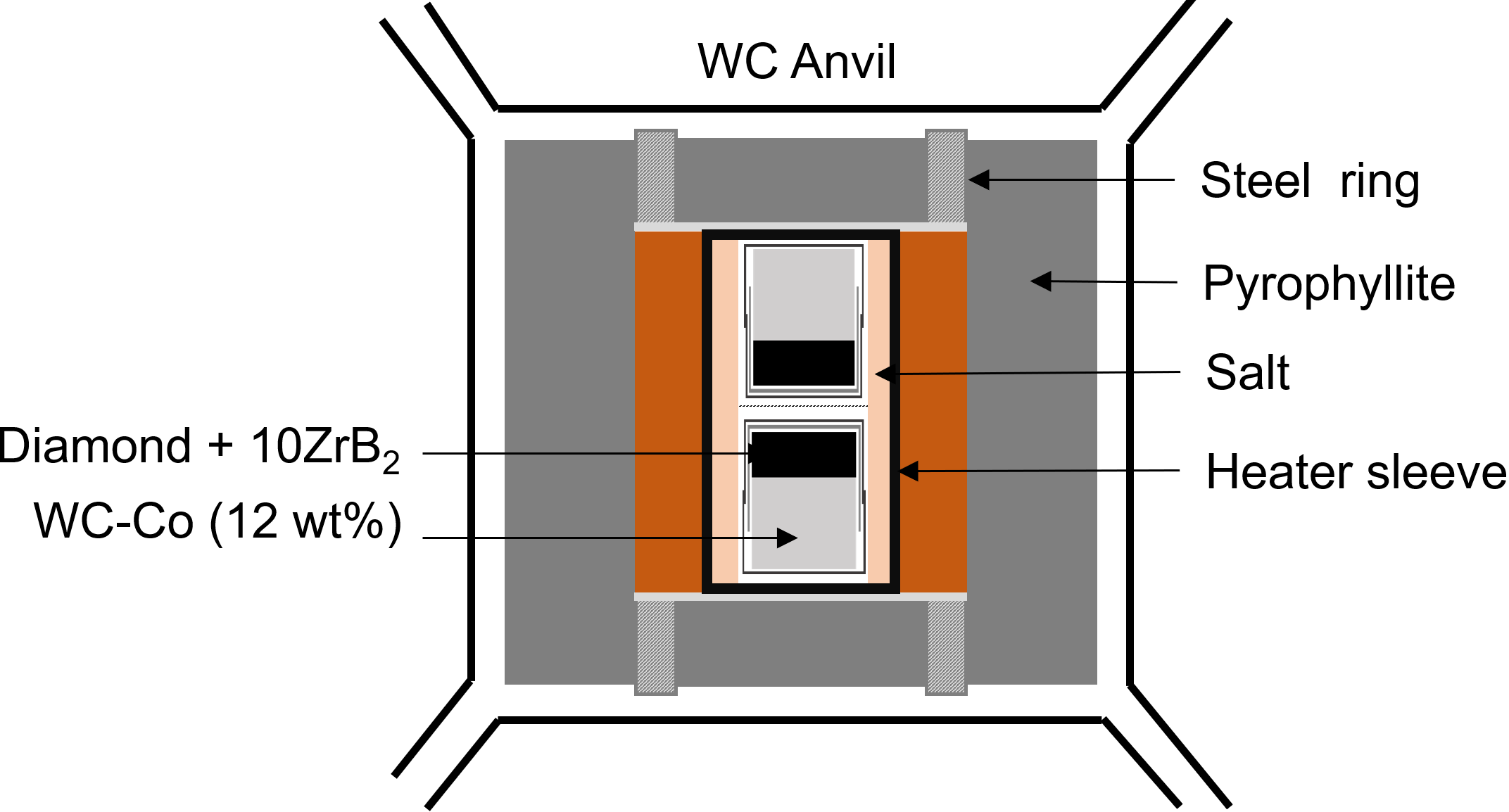}
  \end{center}
  \caption{Schematic diagram of the cubic press capsule used for sintering}
  \label{fig:HPHTCapsule}
  \end{figure}

The cylindrical tool ($\phi$ = 16 mm; height = 12 mm) consisted of a 2 mm thick polycrystalline diamond layer sintered on top of a WC-Co (12 wt\%) substrate. Diamond powder (mean particle size: \SI{16}{\micro m}) with 10 wt\% ZrB\textsubscript{2} additions (\emph{ZrB\textsubscript{2}-PCD}) and without (\emph{STD-PCD}) were sintered at $\approx$ \SI{7.0}{GPa} and \SI{1700}{\celsius} in a cubic press system.  Figure \ref{fig:HPHTCapsule} illustrates the assembly used for synthesis.

Samples 10 mm in diameter and 1 mm thick were removed from the PCD layer using electro-discharge machining for the powder X-ray diffraction (PXRD) analysis. Cross-sectional samples were cut through the diamond layer into the WC-Co substrate and polished to a \SI{1}{\mu m} finish using a diamond impregnated resin wheel for the scanning electron microscopy (SEM) analyses. 
Phase analysis of the as-sintered samples was performed using a Bruker D2 Phaser diffractometer equipped with a Co x-ray source, Lynx eye PSD detector, with pseudo Bragg-Brentano geometry. Polished cross-sections were imaged using  backscattered electrons (BSE) in an JEOL7001F SEM (10 keV) equipped with energy dispersive spectroscopy (EDS) capabilities. Image segmentation and quantitative microstructural analyses were performed on the BSE-SEM images using the MIPAR software package \citep{sosa2014} in order to determine diamond phase fraction, binder phase fraction, diamond grain size, and diamond-phase contiguity \citep{german1985}.
Thin lamellae were removed from the binder phases of the two sintered samples using focused (Ga) ion-beam (FIB)-SEM. The thin sections were then analysed using transmission Kikuchi diffraction (TKD)-EDS in the SEM and selected area electron diffraction (SAED) in the transmission electron microscope to determine the crystal structure of the binder phases. In-situ PXRD data were collected on a PANalytical X\'Pert Pro diffractometer equipped with an Anton Paar HTK1200 reaction chamber ($\lambda_{Co}$ = \SI{1.789}{\angstrom}). The non-ambient heating was performed under vacuum, at a pressure of \SI{0.1e5}{\kPa} using a stepped profile from \SI{800}{\celsius} up to \SI{1100}{\celsius}. PXRD patterns were collected every 15 minutes from 15-65 $2 \theta \degree$. The collected PXRD data were processed using Bruker AXS TOPAS (Version 5, 2014) to track the microstructural evolution using the whole pattern fitting approach. The patterns collected at non-ambient conditions were sequentially refined to track the phase fractions and lattice parameters of the crystallographic phases.
A probe corrected JEOL ARM 200F transmission electron microscope operated at 200 kV was used for the analysis in the present study. High angle annular dark field (HAADF) scanning transmission electron microscopy (STEM) imaging sensitive to atomic number variations within the sample, and electron energy loss spectroscopy (EELS) spectrum imaging (SI) were performed using the DualEELS mode on a Gatan GIF Quantum ERS spectrometer. 
Carbon (diamond and graphite) and cobalt phase maps were extracted by performing multi linear least squares fitting (MLLS) on the 3D data cube after splicing the low-loss and high-loss SIs together. Reference spectra for the background, diamond, graphite, and cobalt phases were extracted from the experimental data set and used in the MLLS analysis.
The thin lamellae were mounted on a single-tilt DENSsolutions heating chip. The samples were heated from \SI{450}-\SI{900}{\celsius} in intervals of \SI{100}{\celsius}. The samples were held for 10 minutes at each temperature interval after which imaging was performed using HAADF-STEM. The presence of suspected graphitic phases were confirmed using STEM-EELS analysis, while elemental analysis of carbide phases were done using STEM-EDS. 

Mechanical testing was performed using a vertical turning lathe (VTL) testing. The PCD tools were brazed into steel holders and tested on a South African Granite (UCS: \SI{200}{MPa}) work-piece rotating at 100 RPM. Testing was performed with a water coolant (wet-VTL) and without (dry-VTL), with each test having slightly different testing parameters. The wet-VTL testing conditions prevented excessive heating on the cutting edge, thus measuring the abrasion resistance of the tool. The tools were tested for 30 passes in the wet-VTL test, thereafter the wear scar area was measured using an optical profilometer. The dry-VTL testing used no coolant, thereby generating more frictional heating at the cutting surface. The dry testing conditions give an indication of the thermo-mechanical wear behaviour of the tools. During dry-VTL testing, the tools were tested at intervals of 2 passes and stopped once the wear scar area progressed into the carbide layer.

\section{Results}
\begin{figure}[h]
  \begin{center}
    \includegraphics[width=\textwidth]{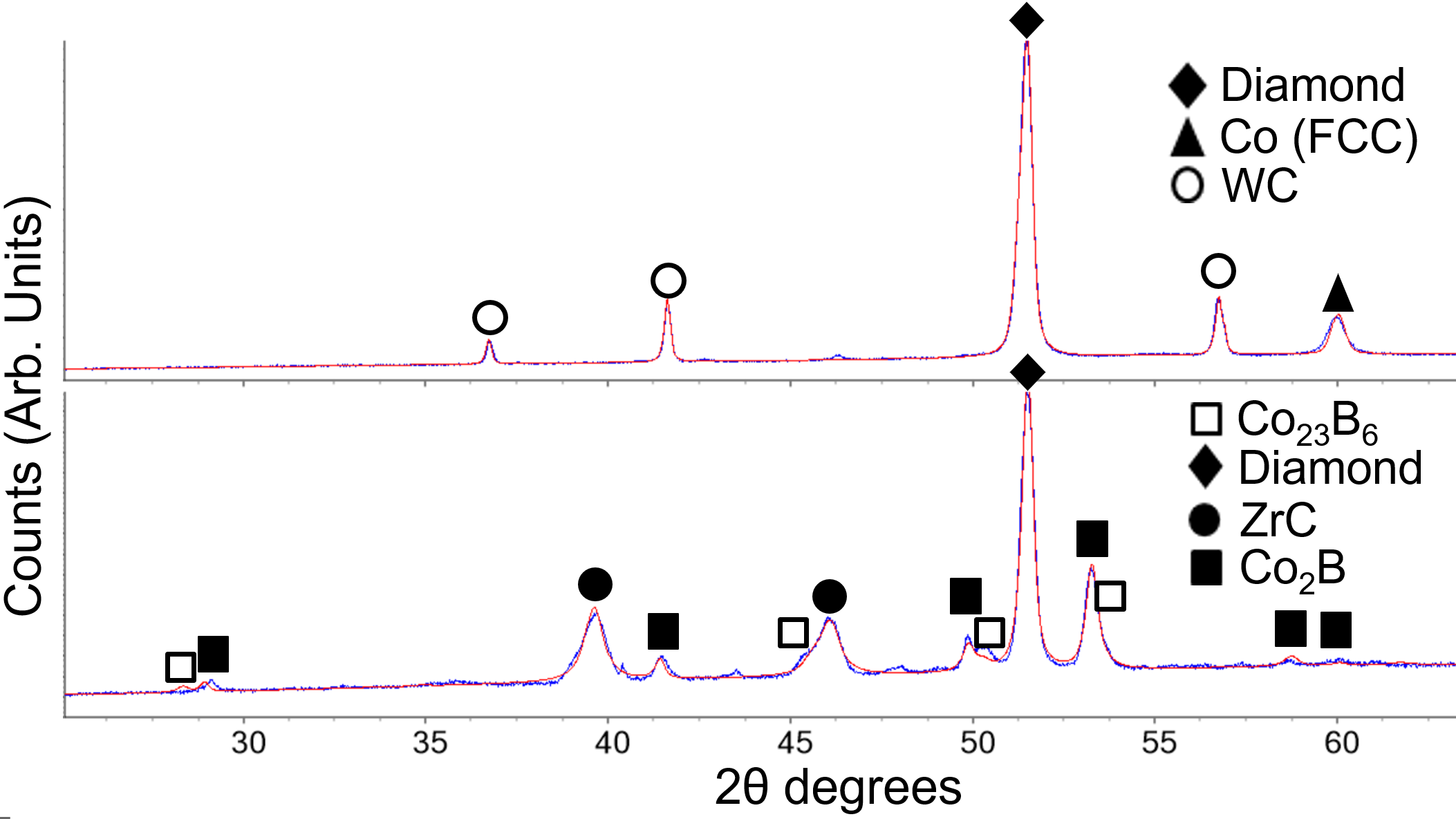}
  \end{center}
  \caption{PXRD analysis of the \emph{ZrB\textsubscript{2}-PCD} and \emph{STD-PCD} starting states.}
  \label{fig:XRDStarting}
\end{figure}

The PXRD phase analyses performed on the starting states (Figure \ref{fig:XRDStarting}) shows that \emph{ZrB\textsubscript{2}-PCD} consisted of diamond, WC, ZrC, Co\textsubscript{2}B (ICSD: 5924), and B\textsubscript{6}Co\textsubscript{23} (ICSD: 613389) phases. The cobalt peak (200) observed in \emph{STD-PCD} was not detected in \emph{ZrB\textsubscript{2}-PCD}, suggesting that either the cobalt concentration fell below the detection limit of the instrument, or that it was not present as a fcc cobalt phase, but rather as cobalt boride phases (Co\textsubscript{2}B and B\textsubscript{6}Co\textsubscript{23}).
 
\begin{figure}[h]
   \begin{center}
     \includegraphics[width=\textwidth]{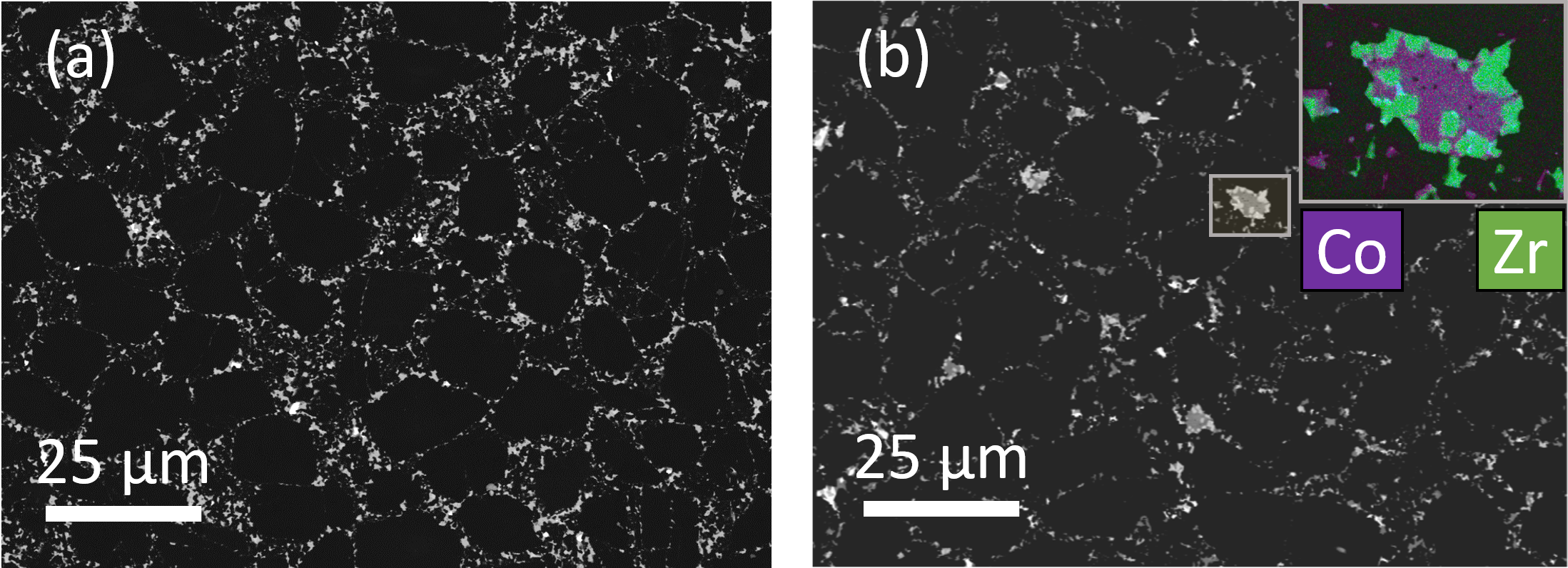}
   \end{center}
   \caption{BSE images of (a) \emph{STD-PCD} and (b) \emph{ZrB\textsubscript{2}-PCD}. Insert shows an EDS map of the larger ZrC-(CoB\textsubscript{2}, B\textsubscript{6}Co\textsubscript{23}) binder pools.}
   \label{fig:BSE_EDS}
\end{figure}

The \emph{STD-PCD} microstructure (Figure \ref{fig:BSE_EDS}a) consisted of diamond particles (dark phase) surrounded by a cobalt binder phase. WC particles (bright phase) can be seen dispersed within the microstructure. The larger diamond particles show good intergrowth with a fine distribution of cobalt phase around the perimeter of the diamond particles. The \emph{ZrB\textsubscript{2}-PCD} sample (Figure \ref{fig:BSE_EDS}b) displayed similar microstructural characteristics, but it also contained several larger ($>$\SI{5}{\mu m}) binder phase regions consisting of a Zr-containing rim and a Co-containing core (Figure \ref{fig:BSE_EDS} coloured EDS map insert).   

\begin{figure}
   \begin{center}
     \includegraphics[width=\textwidth]{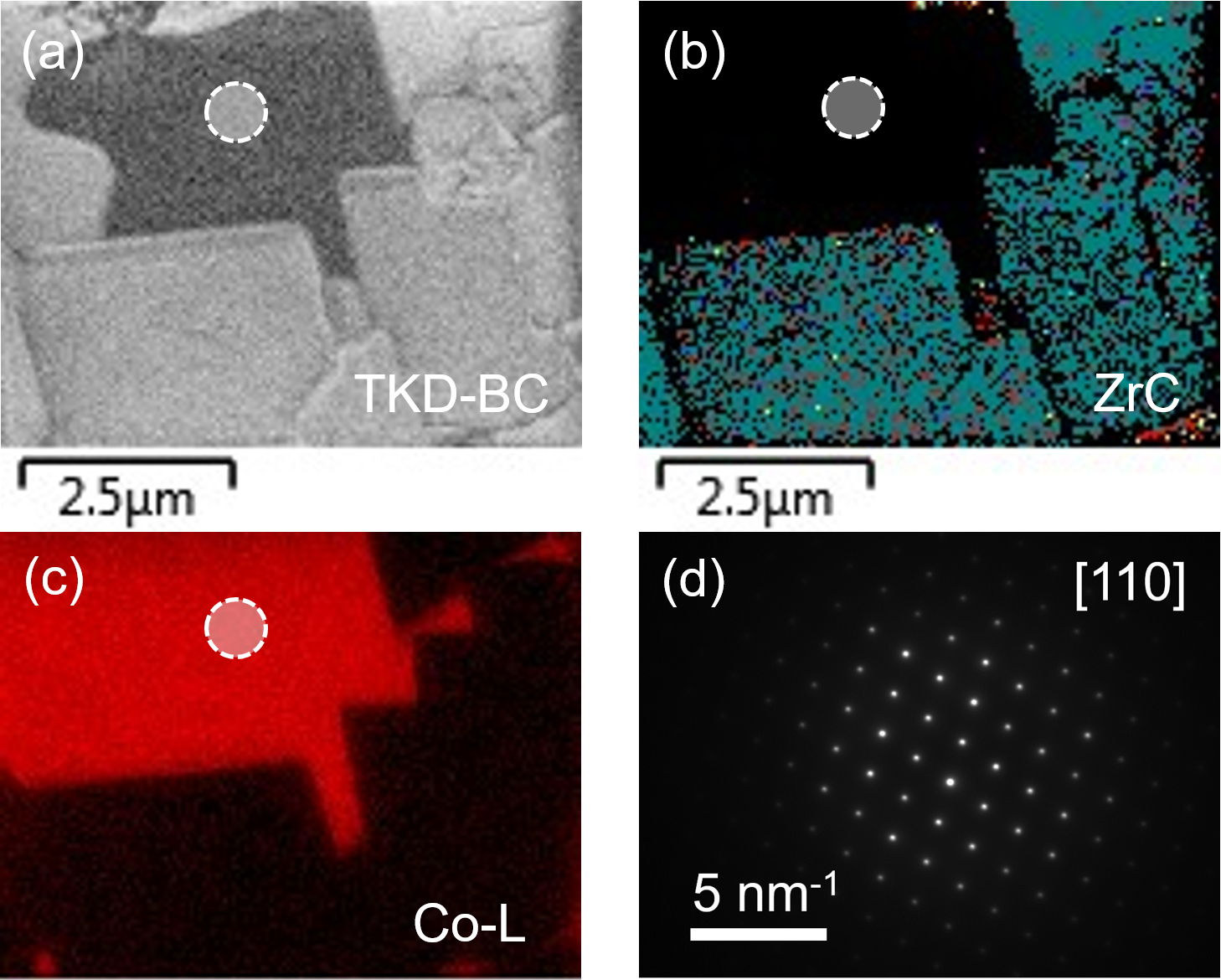}
   \end{center}
   \caption{TKD-EDS analysis of the large binder phase particle showing (a) band contrast, (b) phase map, (c) Co-L EDX map, and (d) B=110 SAED the of B\textsubscript{6}Co\textsubscript{23} phase.}  
   \label{fig:TKD_SAED}
\end{figure}

TKD-EDS crystallographic phase and elemental analysis confirmed that the larger binder phase area consisted of a ZrC (rim) and an inner (non-indexed) Co-containing phase (Figure \ref{fig:TKD_SAED}). Experimental SAED patterns of this Co-containing phase collected at three different zone axes orientations were successfully matched to the [110], [213], and [112] simulated diffraction patterns of the B\textsubscript{6}Co\textsubscript{23} (ICSD: 613389) phase. This result is in agreement with the bulk PXRD analysis that suggests that boron from the ZrB\textsubscript{2} additions reacted with the infiltrated cobalt metal to form boride (Co\textsubscript{2}B and B\textsubscript{6}Co\textsubscript{23}) phases, while the zirconium reacted with carbon to form ZrC particles during sintering. 
 
Quantitative image analysis were performed on BSE images taken at several locations on the polished cross-sectional samples (Table \ref{table:microstructure}). The \emph{ZrB\textsubscript{2}-PCD} sample had a lower diamond content (90.3 vol. \%) compared to the \emph{STD-PCD} sample due to the presence of an additional ~4.5 vol.\% ZrC particles in the microstructure. Consequently, the diamond-diamond contact area (contiguity) was also lower (55 \%) for the \emph{ZrB\textsubscript{2}-PCD} sample due to the presence of this third phase.   

\begin{table}[!ht] 
  \centering
  \caption{Quantitative image analysis for the as-sintered samples.}
  \label{table:microstructure}
  \smaller
  \begin{tabular}{l c c c}
    \toprule
    \textbf{Sample} & \textbf{Diamond ($f_v$)}& \textbf{Cobalt ($f_v$)} & \textbf{Contiguity (\%)} \\
    \midrule
    \emph{STD-PCD} & 93.1 (0.4) & 6.9 (0.4) & 61 (2) \\
    \emph{ZrB\textsubscript{2}-PCD} & 90.2 (0.5) & 4.5 (0.1) & 55 (4) \\
    \bottomrule
  \end{tabular}
\end{table}

 \subsection{In-situ annealing of polycrystalline diamond}
\begin{figure}
   \begin{center}
     \includegraphics[width=\textwidth]{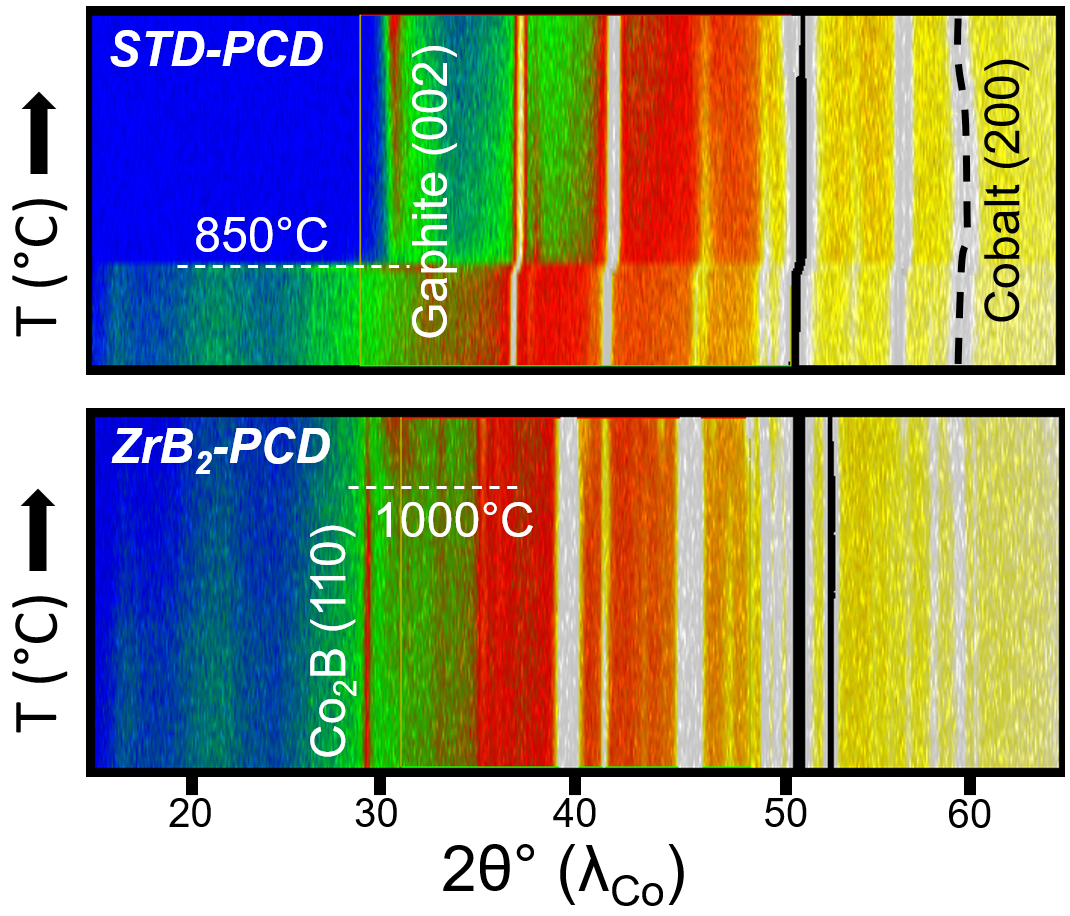}
   \end{center}
   \caption{Topographical plot of the in-situ PXRD intensity showing the microstructural changes that occurred in the samples during high temperature annealing in vacuum.}
   \label{fig:INSITU_XRD}
  \end{figure}

\begin{figure}
   \begin{center}
     \includegraphics[width=\textwidth]{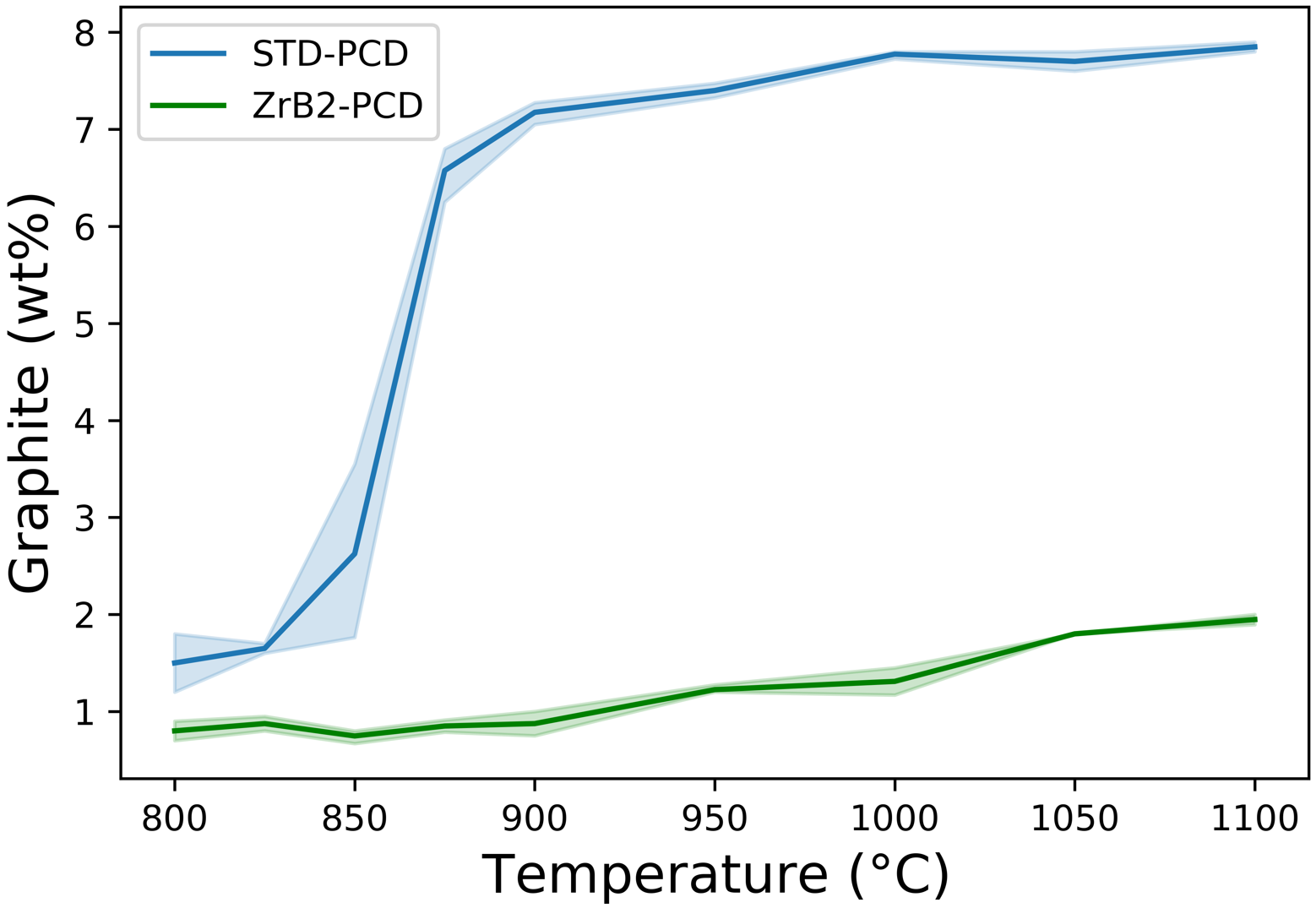}
   \end{center}
   \caption{Quantitative phase analysis obtained from the PXRD data for graphite in the vacuum annealed samples.}
   \label{fig:GraphiteQuant}
  \end{figure}

 The topographical view of the non-ambient PXRD analysis of the \emph{STD-PCD} and \emph{ZrB\textsubscript{2}-PCD} samples are shown in Figure \ref{fig:INSITU_XRD}. The \emph{STD-PCD} showed a decrease in the cobalt lattice parameter accompanied by the formation of Co\textsubscript{3}W\textsubscript{3}C $\eta$-phase for temperatures between 800 - \SI{850}{\celsius}, which is consistent with previous high-temperature annealing studies of PCD \citep{westraadt2015a, scott2019}. According to the PXRD data, graphite was first observed at \SI{850}{\celsius} in the \emph{STD-PCD} sample. Quantitative phase analysis of graphite for the \emph{STD-PCD} sample (Figure \ref{fig:GraphiteQuant}) shows a rapid increase of the graphite phase fraction at \SI{850}{\celsius}, which then plateaued at approximately 7 wt\%. There were several additional microstructural changes that occurred in the \emph{STD-PCD} sample during high-temperature annealing in vacuum, but this will be the subject of future study.   

 In contrast, the \emph{ZrB\textsubscript{2}-PCD} sample showed no observable microstructural changes during vacuum annealing at 800 - \SI{1000}{\celsius}. According to the PXRD data,graphite was first observed after \SI{1000}{\celsius} in the \emph{ZrB\textsubscript{2}-PCD} sample. The peak heights of the Co\textsubscript{2}B phase decreased after \SI{1000}{\celsius} heat treatment. Quantitative phase analysis of graphite for the \emph{ZrB\textsubscript{2}-PCD} sample (Figure \ref{fig:GraphiteQuant}) showed a gradual increase, which remained below 2 wt\% up until \SI{1100}{\celsius}. The decrease in the Co\textsubscript{2}B phase fraction could not be accurately quantified due to its low relative abundance in the presence of other highly crystalline phases as well as a loss in the quality of the data which is affected by additional scattering effects brought about at elevated temperatures. Therefore, the possible decomposition of the boride phases into a pure cobalt phase could not be confirmed from this particular PXRD data set.     

\begin{figure}
  \begin{center}
   \includegraphics[width=\textwidth]{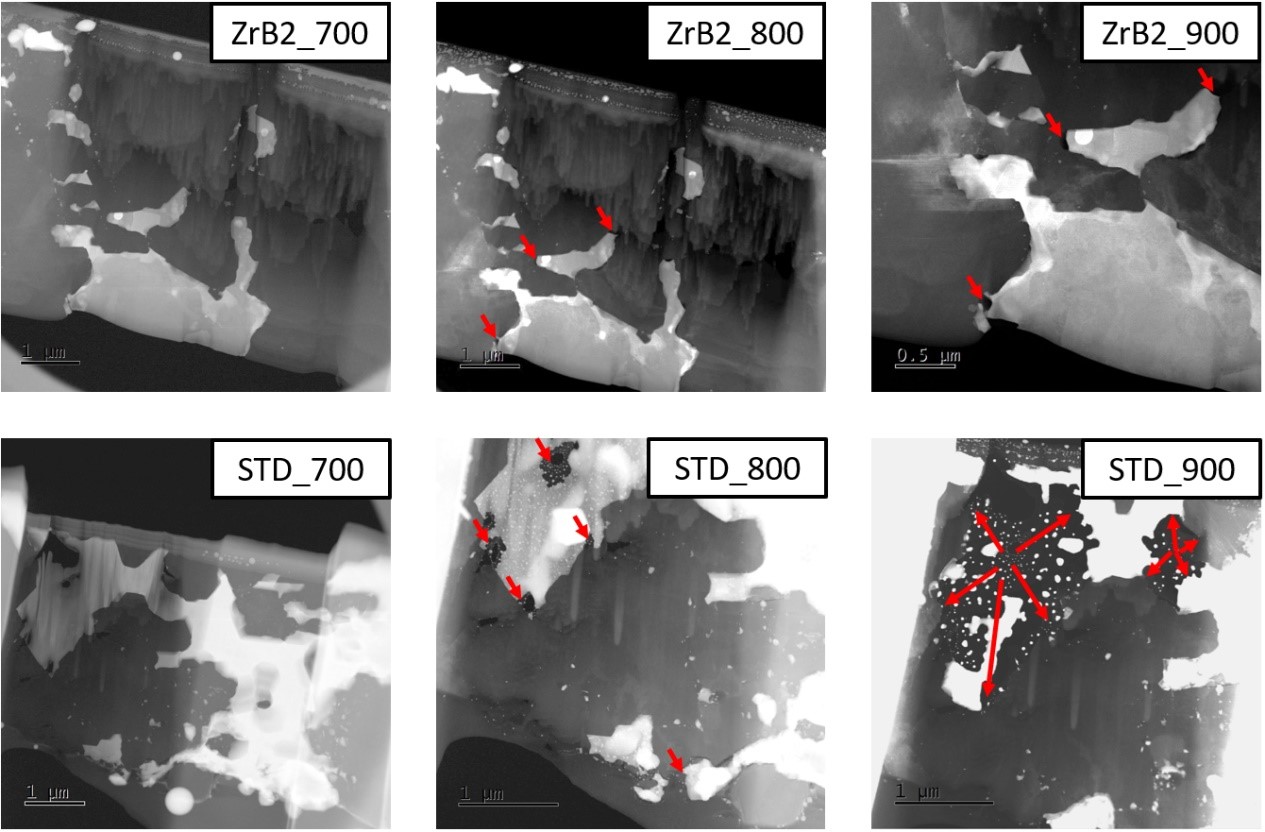}
  \end{center}
  \caption{HAADF-STEM images of the \emph{ZrB\textsubscript{2}-PCD} (top) and \emph{STD-PCD} (bottom) samples after 10 minutes of in-situ TEM annealing at the various temperatures. The confirmed presence of graphitic carbon is indicated with red arrows.}
 \label{fig:HEAT_STEM}
 \end{figure}

\begin{figure}  
  \begin{center}
   \includegraphics[width=\textwidth]{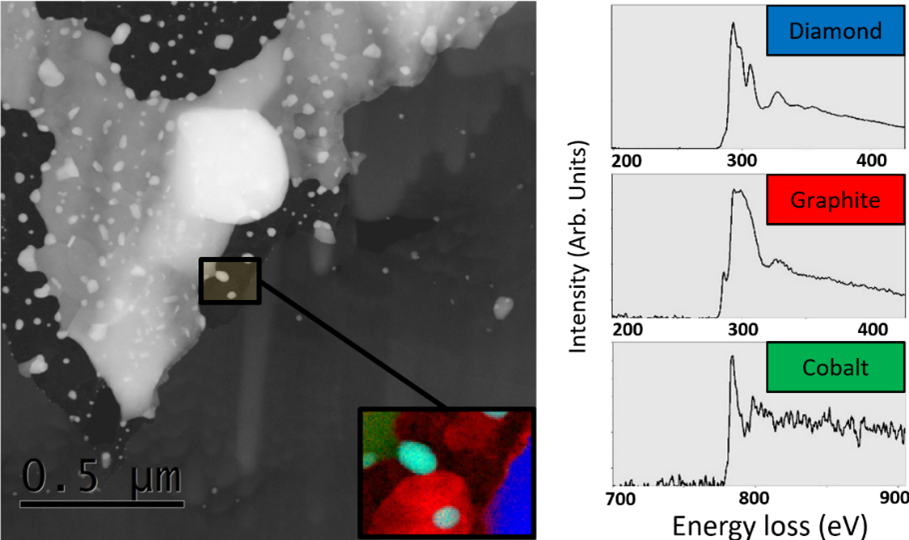}
  \end{center}
  \caption{HAADF-STEM image of a Co-W binder pool after in-situ TEM annealing at \SI{850}{\celsius} for 10 minutes. The insert shows a STEM-EELS coloured phase map of the binder/diamond interface, indicating the distribution of diamond (blue), graphite (red) and cobalt (green). The corresponding electron energy loss spectra used in the MLLS fitting are shown on the right.}
  \label{fig:STEM_EELS_STD}
 \end{figure}

 HAADF-STEM images of the samples after in-situ TEM annealing for 10 minutes at \SI{700}{\celsius}, \SI{800}{\celsius}, and \SI{900}{\celsius} are shown in Figure \ref{fig:HEAT_STEM}. The samples were evaluated for suspected graphite using image contrast which was then confirmed using STEM-EELS analysis, indicated by red arrows on the HAADF-STEM images. No graphite formation was observed for annealing temperatures up to \SI{700}{\celsius}. The onset of graphite formation occurred within 10 minutes of heating at \SI{800}{\celsius} for the \emph{STD-PCD} sample, with multiple graphitic carbon sites confirmed at the cobalt/diamond interface. Graphite formation continued rapidly in the \emph{STD-PCD} sample as the temperature was increased to \SI{900}{\celsius}, extending into the cobalt binder pool. Figure \ref{fig:STEM_EELS_STD} shows a higher magnification HAADF-STEM image and a coloured phase map insert of the cobalt-diamond interface after \SI{800}{\celsius}  annealing, thus confirming the presence of graphitic carbon.      
 In comparison, HAADF-STEM imaging of the in-situ TEM annealed \emph{ZrB\textsubscript{2}-PCD} sample showed only minor signs of graphite formation after 10 minutes of heating at \SI{800}{\celsius}. Subsequent annealing at \SI{900}{\celsius} for 10 minutes, showed no additional graphitic nucleation sites and no appreciable growth of the existing graphite sites.

 \subsection{Thermo-mechanical VTL testing}
\begin{figure}
  \begin{center}
   \includegraphics[width=\textwidth]{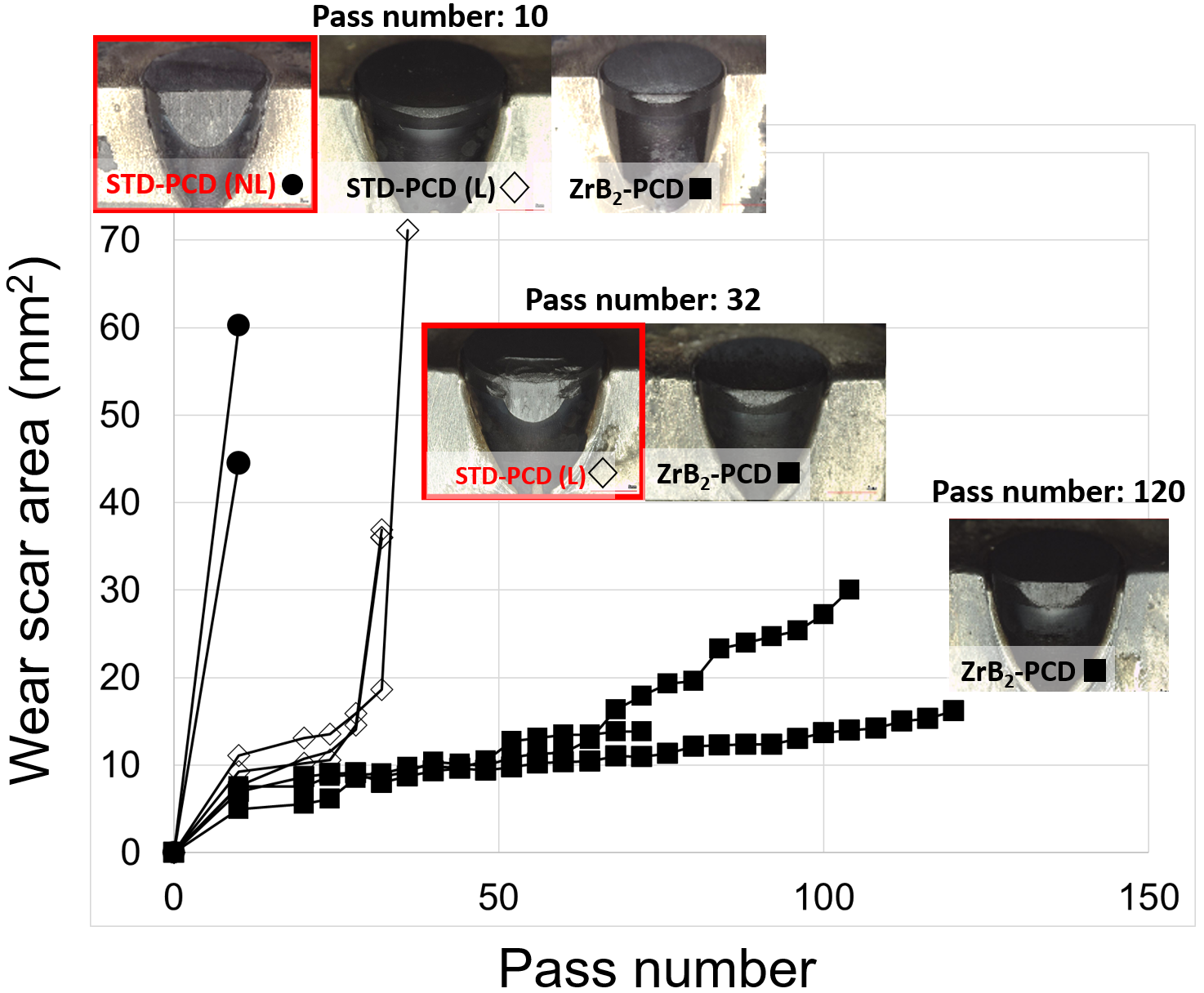}
  \end{center}
  \caption{Thermo-mechanical wear testing of the samples using the dry-VTL testing conditions. (\textbf{NL}: non-leached; \textbf{L}: \SI{250}{\mu m} acid leach)}
  \label{fig:DRY_VLT}
 \end{figure}

 Vertical turning lathe (VTL) testing were performed to investigate the thermo-mechanical wear behaviour of the polycrystalline diamond materials sintered with ZrB\textsubscript{2} additions. 
 The dry-VTL testing results are shown in Figure \ref{fig:DRY_VLT}. The \emph{STD-PCD} sample tools (n=2) (\emph{STD-PCD (NL)}) failed after the first set of 10 passes after developing a large wear scar that extended into the WC-Co substrate. \emph{STD-PCD} tools (n=3) leached to a depth of \SI{250}{\mu m} (\emph{STD-PCD (L)}) showed initial wear scar progressions similar to the \emph{ZrB\textsubscript{2}-PCD} samples, but failed catastrophically after 30-32 passes. The \emph{ZrB\textsubscript{2}-PCD} tools (n=3) had a steady wear scar progression until 75, 104, and 120 passes, when the wear scars reached the WC-Co substrate and the testing was stopped. 

\begin{figure}
  \begin{center}
    \includegraphics[width=\textwidth]{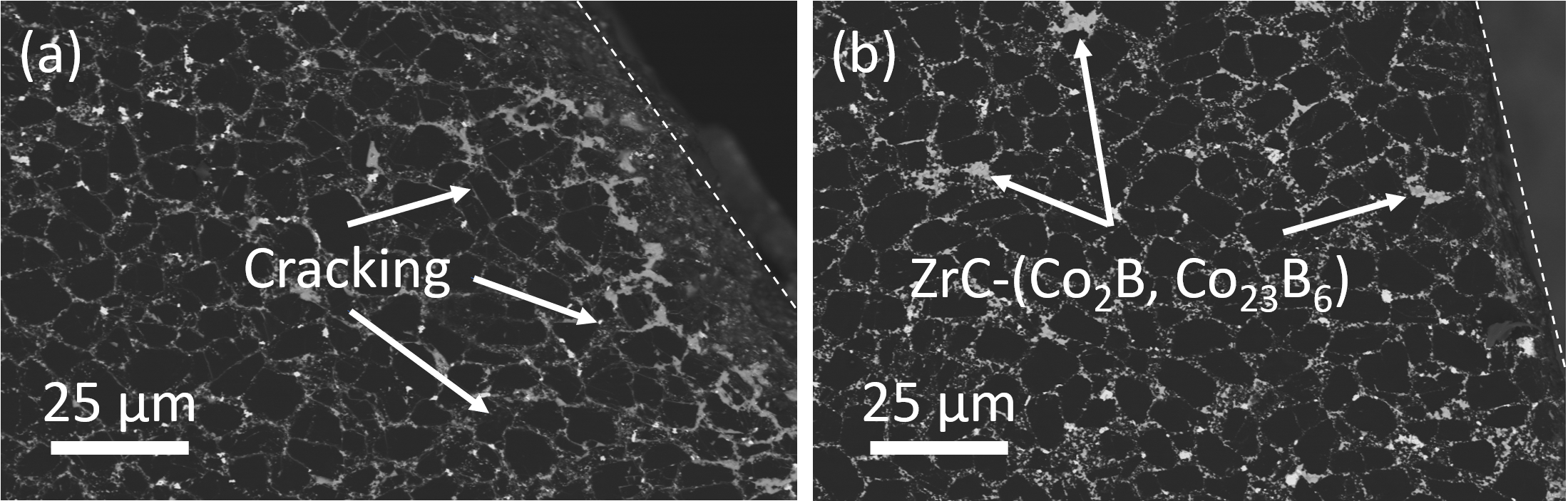}
  \end{center}
  \caption{BSE-SEM images of the wear scar area for (a) \emph{STD-PCD} and (b) \emph{ZrB\textsubscript{2}-PCD} following the dry-VTL testing.}
  \label{fig:WEAR_SEM}
 \end{figure}

 Following the dry-VTL testing, SEM analysis of the wear scar area was performed in order to investigate the failure mechanisms for both materials. Figure \ref{fig:WEAR_SEM} shows the BSE-SEM images of the tools cross-sectioned through the wear scar area indicated by the dashed line. The \emph{STD-PCD} sample showed a compromised diamond structure with transgranular cracks and several large SiO\textsubscript{2} inclusions (confirmed with EDS analysis - not shown). The origin on the SiO\textsubscript{2} inclusions is probably due to contaminants from the polishing wheel being deposited into larger cavities created during the high-temperature wear test. The premature failure of the tool (10 passes) was due to thermal degradation for the unleached \emph{STD-PCD} sample.    

 The \emph{ZrB\textsubscript{2}-PCD} wear scar did not show extensive transgranular cracking observed in the \emph{STD-PCD} sample and there were no SiO\textsubscript{2} inclusions embedded during the polishing process. This indicates that the integrity of the diamond structure was maintained through the duration of the high-temperature wear test. There were no evidence for thermal degradation in the \emph{ZrB\textsubscript{2}-PCD} sample from the SEM analysis, which confirms that the ZrB\textsubscript{2} additions improved the thermal stability of the PCD tool.  
 
 The truncated wet-VTL testing generate lower temperatures at the tool interface and is a measure of the tools' ability to withstand abrasive wear. The mechanical properties in PCD materials are related to the diamond-diamond contact area (Contiguity \%) and the strength of the diamond-diamond intergrown structure. The \emph{ZrB\textsubscript{2}-PCD} tools (n=4) showed a larger wear scar ($17.2 \pm 0.3$ mm\textsuperscript{2}) compared to the \emph{STD-PCD} ($10.1 \pm 0.4$ mm\textsuperscript{2}) tools (n=4) for the wet-VTL testing truncated to 30 passes. This result implies the abrasion wear of the PCD tools was reduced by the addition of ZrB\textsubscript{2} to the diamond powder.

\section{Discussion}

 \subsection{Effect on the as-sintered microstructure}
 The addition of ZrB\textsubscript{2} to the diamond powder prior to sintering resulted in an intergrown diamond microstructure containing residual cobalt-boride (Co\textsubscript{2}B, Co\textsubscript{23}B\textsubscript{6}) phases. Larger binder areas consisting of a ZrC outer rim and an cobalt-boride inner core are distributed throughout the microstructure (Figure \ref{fig:BSE_EDS} and Figure \ref{fig:WEAR_SEM}). The ZrB\textsubscript{2} phase reacted with the molten cobalt during sintering to form the cobalt-boride and ZrC phases. The diamond phase fraction ($f_v$) and diamond phase contiguity (\%) of the \emph{ZrB\textsubscript{2}-PCD} was lower than the \emph{STD-PCD} material due to the formation of a ZrC secondary phase. PXRD analysis could not detect a pure cobalt phase in the \emph{ZrB\textsubscript{2}-PCD} sample and all the cobalt were converted into cobalt-boride phases.  

 \subsection{Effect on the catalytic graphitisation}
 Metal-boride interlayers are commonly used to prevent catalytic graphitisation of diamond coatings grown on WC-Co and steel cutting tools \citep{tang2002, damm2017}. Thus the catalytic conversion of diamond to graphite for the cobalt-boride binder was expected to be lower relative to the pure cobalt binder phase present in the \emph{STD-PCD} sample. The \emph{ZrB\textsubscript{2}-PCD} sample only showed measurable bulk graphitisation at temperatures above \SI{1000}{\celsius}, compared to the \emph{STD-PCD} sample that graphitised at \SI{850}{\celsius} during the in-situ PXRD analysis. During in-situ TEM annealing, STEM-EELS analysis showed that graphite was first observed at \SI{800}{\celsius} for both samples, but the subsequent graphite growth was more pronounced for the \emph{STD-PCD} sample at \SI{900}{\celsius}. This shows that the ZrB\textsubscript{2} additives resulted in a more thermally stable (Co\textsubscript{2}B and B\textsubscript{6}Co\textsubscript{23}) binder phase with a lower propensity for diamond-to-graphite conversion during elevated temperature annealing performed in vacuum. 

 \subsection{Effect on the thermo-mechanical wear properties}
 The \emph{ZrB\textsubscript{2}-PCD} sample showed a significant improvement in the high temperature dry-VTL testing as compared to the \emph{STD-PCD}. The high temperatures generated at the cutting interface resulted in transgranular cracking and cavity formation, which resulted in early failure of the unleached \emph{STD-PCD} tools after just 10 passes. The thermally degraded microstructure of the \emph{STD-PCD} tool was similar to that of a previous study on high temperature milling of granite using a similar PCD tool \citep{westraadt2015a}.  The leached (\SI{250}{\mu m} depth) \emph{STD-PCD} sample showed a wear scar progression similar to the \emph{ZrB\textsubscript{2}-PCD} sample for the initial stages, but then rapidly degraded and failed catastrophically after 32 passes. In contrast the \emph{ZrB\textsubscript{2}-PCD} sample showed no obvious microstructural evidence of thermal degradation and exhibited a controlled wear scar progression into the WC-Co substrate after 120 passes. 

 The abrasive wear properties and hardness of PCD tools are directly related to the intergrown diamond network, where a higher phase fraction of diamond have been linked to an increased hardness and "grinding resistance" for PCD materials sintered with less binder phase at elevated pressures \citep{akaishi1988}. Ultrahard ceramic materials such as PCD have an inherently low fracture toughness (\SI{8}{\mega Pa {m^{0.5}}}), which is then further negatively affected by microstructural flaws \citep{scott2017}. PCD materials containing poorly bonded diamond-diamond particles, large inclusions, or cavities post-sintering would negatively affect the fracture behaviour of the tool during abrasive testing. The \emph{ZrB\textsubscript{2}-PCD} sample had a larger wear scar compared to the \emph{STD-PCD} sample in the truncated wet-VTL test. The reduced abrasive wear resistance of \emph{ZrB\textsubscript{2}-PCD} sample could be due to the larger binder pools with ZrC inclusions, which resulted lower diamond phase fraction, reduced diamond-diamond intergrowth, and could also have acted as microstructural flaws negatively affecting the fracture behaviour of the tool. In addition, boron could have influenced the ability of cobalt to act as an effective sintering aid for diamond intergrowth.     

  Since the primary advantage of PCD tools in application is its high abrasion resistance, it is imperative that its inherent mechanical performance should not be compromised by the additive. The VTL testing data highlights an important trade-off that exist when evaluating new thermally stable PCD materials based on binder additives. This study demonstrate the intimate link between processing-microstructure-properties, and highlights the importance of tailoring the processing and/or microstructure to achieve the desired balance between abrasive wear resistance and thermal stability.  

  In order to improve the abrasion resistance of the \emph{ZrB\textsubscript{2}-PCD} material, while still retaining its exceptional thermal stability a number of optimisations can be explored. A reduction in the particle size and phase fraction of the residual ZrC phase is required. This could be achieved via better control of the starting particle size, additive content, or more homogeneous mixing of the ZrB\textsubscript{2} additives. Furthermore, increasing the sintering conditions using higher pressure and/or higher temperature could potentially offset the negative effects that boron could have on the cobalt to act as a liquid phase sintering aid. 

\section{Conclusions}
From the characterisation data presented and discussed here we can draw the following conclusions about the materials:
 \begin{itemize}
  \item The addition of 10 wt\% ZrB\textsubscript{2} to the diamond powder prior to sintering yielded an intergrown diamond structure with a complete conversion of the residual binder to cobalt-boride (Co\textsubscript{2}B, Co\textsubscript{23}B\textsubscript{6}) phases.
  \item The onset of bulk graphitisation in the \emph{ZrB\textsubscript{2}-PCD} material was improved to temperatures above \SI{1000}{\celsius} compared to the \SI{850}{\celsius} onset observed for the \emph{STD-PCD} material.
  \item The \emph{ZrB\textsubscript{2}-PCD} tool had excellent thermo-mechanical wear performance in dry-VTL testing and the microstructure showed no evidence of thermal degradation. 
  \item However, the abrasion wear performance of the \emph{ZrB\textsubscript{2}-PCD} tool in wet-VTL testing was compromised, due to the presence of 4.5 vol.\% ZrC phase, resulting in reduced diamond-diamond intergrowth (diamond $f_v$ and contiguity \%). 
 \end{itemize}
\newpage

\section*{Data availability}
The raw data required to reproduce these finding are available to download after a request to the corresponding author. 

\section*{Declaration of Competing Interest}
The authors declare that they have no known competing financial interests or personal relationships that could have appeared to influence the work reported in this paper. 

\section*{\emph{Acknowledgement}}
JEW and MJ gratefully acknowledge financial support of the National Research Foundation of South Africa (Grant number 70724). We also acknowledge the assistance of Mohsen Danaie and Christopher S. Allen from the electron Physical Science Imaging Centre (ePSIC) at the Diamond Light Source (Ltd) with the collection of the in-situ TEM data (Instrument E01, proposal MG26338) and David Aldmington from the Global Innovation Centre (GIC) at Element Six (Pty) Ltd with the collection of the in-situ PXRD data.   

\bibliography{PCD106_P1}

\end{document}